\documentclass[10pt,letterpaper,twocolumn]{article} 
\usepackage{ol2}
\usepackage[draft]{hyperref}
\usepackage{amsmath}
\newcommand{\be}{\begin{equation}}
\newcommand{\ee}{\end{equation}}
\newcommand{\bea}{\begin{eqnarray}}
\newcommand{\eea}{\end{eqnarray}}
\newcommand{\bes}{\begin{subequations}}
\newcommand{\ees}{\end{subequations}}
\newcommand{\PT}{\mathcal{PT}}

\begin{document}
\twocolumn[ 
\title{Tunable nonlinear $\PT$-symmetric defect modes with an atomic cell}
%
%
\author{Chao Hang$^{1}$, Dmitry A. Zezyulin$^{2}$, Vladimir V. Konotop$^{2}$, and Guoxiang Huang$^{1}$}
\address{
$^1$State Key Laboratory of Precision Spectroscopy
and Department of Physics, East China Normal University, Shanghai
200062, China \\
$^2$Centro de F\'isica Te\'orica e Computacional and Departamento
de F\'isica, Faculdade de Ci\^encias, Universidade de Lisboa,
Instituto para Investiga\c{c}\~ao Interdisciplinar, Avenida
Professor Gama Pinto 2, Lisboa 1649-003, Portugal }
\begin{abstract}  We propose a scheme of creating  a tunable highly nonlinear
defect in a one-dimensional photonic crystal.  The defect consists of an atomic cell filled in with two isotopes of three-level atoms.  The
probe-field refractive index of the defect can be made parity-time ($\PT$) symmetric, which is achieved
by proper combination
of a control field and of Stark shifts induced by a
far-off-resonance field. In the $\PT$-symmetric   system families of stable nonlinear defect modes can be formed by the probe field.
\end{abstract}

\ocis{190.0190,190.6135,190.3270.}

] 

\noindent


Defects are known to play an important role in controlling,
manipulating and guiding light. Depending on the physical nature of a defect,  one can
distinguish conservative and nonconservative (i.e. active or
dissipative) defects, as well as linear and nonlinear ones,
which display  very different properties with respect to the
modes they support. In particular, nonlinear conservative defects
embedded in a linear structure can support families of the modes,
whose propagation constant depends on the field intensity
(representing the so called families of solitons). Meantime, the modes supported by
nonconservative (active) defects have been  attracting increasing
recent attention, see \cite{Malomed1,ZKK} and \cite{dissipative} for a brief review. Shapes and propagation constants of such modes are strictly determined  by the system parameters
(mathematically, such modes are isolated fixed points).

For applications it is desirable to have defects
with tunable parameters. To this end there were suggested to
employ the electro-optical effect~\cite{electro} implemented
experimentally using a nematic liquid crystal defect
layer~\cite{liquid1,liquid2}, to control a liquid crystal cell
(defect) by magnetic field~\cite{magnetic}, or to enhance
conservative defect modes through the parametric
resonance~\cite{parametric}. Tunable active defects can be
implemented by using doping of a desired domain of the guiding
medium by active impurities like in the experimental setting of
Ref.~\cite{Ruter}, where a  parity-time ($\PT$-)
symmetric~\cite{Bender} defect, i.e. a structure with the refractive index obeying
$n(x)=\bar{n}(-x)$~\cite{Muga},  was created (hereafter overbars stand for complex conjugation).

In the present Letter, we suggest a way of implementing a tunable {\em
nonlinear} defect in a photonic crystal. The defect can be
transformed from a dissipative to a $\PT$ symmetric simply by changing the external control field. In the $\PT$-symmetric case it  allows for
existence of continuous families of localized guided modes. (We emphasize the difference between our   statement and the recent  studies of defect modes with
a localized defect  placed in a $\PT$-symmetric
lattice~\cite{Hu,WW}).

\begin{figure}
\centering
\includegraphics[width=0.85\columnwidth]{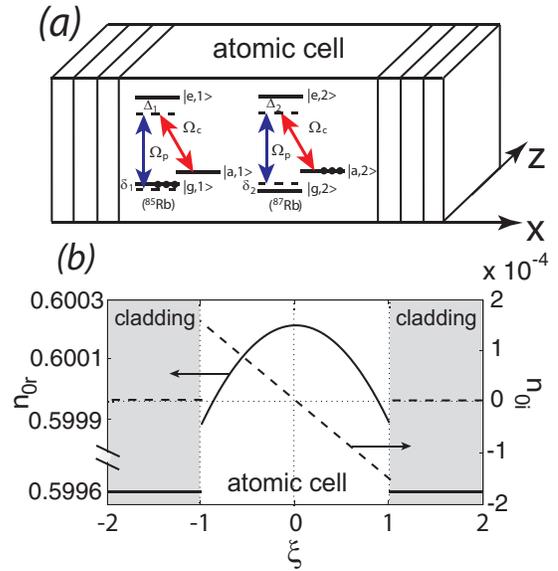}
\caption{(Color online) (a) Geometry of the photonic crystal, the energy-level diagram, and Raman resonance scheme of a binary mixture.
The initially populated levels
are indicated by the black dots. (b) Distribution of real $n_{0r}$ (solid line) and imaginary $n_{0i}$ (dashed line) parts of the linear refractive index induced by the fields (\ref{field}). }
\label{fig1}
\end{figure}

Following~\cite{HHK}, where an optically active
$\PT$-symmetric atomic system was proposed,   we consider a cell in a
one-dimensional photonic crystal which is filled in by  a mixture
of two isotopes ($s=1,2$) of three-level atoms [Fig.~\ref{fig1}(a)].
Like in the original
work~\cite{Brien}, where the system was introduced for creating
large real refractive indexes (see also~\cite{Yavuz1} for
comparison with alternative schemes), each isotope is represented
by a three-level atom with two ground-state states, $|g,s\rangle$
and $|a,s\rangle$, and one excited state, $|e,s\rangle$.
%
%

A weak probe field propagating along $z$-direction
and having the wavenumber $k_{p}$ and amplitude $E_p$, induces
transitions between the states $|g,s\rangle$ and $|e,s\rangle$ with the Rabi frequency
$\Omega_{p}=p_{eg}E_p/\hbar$, while a strong control field of a
wave number $k_{c}$ and amplitude $E_c$ results in coupling
$|a,s\rangle$-$|e,s\rangle$ with the Rabi frequency
$\Omega_c=p_{ea}E_c/\hbar$. Here $p_{eg}$ and $p_{ea}$ represent the electric
dipole matrix elements of the respective inter-level transitions (we assume them to be equal
for the both isotopes, i.e. to be independent on $s$).
The one-  and  two-photon detunings are given respectively by
$\Delta_{s}=\omega_e^{s}-\omega_a^{s}-\omega_c$ and
$\delta_{s}=\omega_a^{s}-\omega_e^{s}-(\omega_p-\omega_c)$, where
$\omega_l^s$ $(l=a, e)$ is the eigenfrequency of the state
$|l, s\rangle$. We also assume that all fields are far-off one-photon
resonance but close to the  two-photon (i.e. Raman) resonance, i.e.
 $\Delta_{s}\gg|\delta_{s}|$. Each $\Lambda-$system is
initially prepared in one of the two ground-state states
[Fig.~\ref{fig1}(a)], such that $s=1$ and $s=2$ isotopes with
$\delta_1>0$ and $\delta_2<0$ provide, respectively, the two-photon
absorption and  gain of the probe field.

Let us  highlight some features of the
defect created by the proposed atomic cell. First, it uses the same mechanism (i.e. the interaction of the laser field with an atomic cell)  to   control   both gain and dissipation.
Second, in resonant atomic media one can create extremely strong
nonlinearities~\cite{AtomicCell} (the generation power of the
$\PT$-symmetric defect modes considered below can be reduced to a
nanowatt range), thus allowing one to explore nonlinear defects and
  to operate with  nonlinear defect modes embedded in a
linear photonic crystal. Third, the shape of the defect can be changed {\em in
situ}.  At the same time, it is also relevant that the possibility of  direct transfer
of the ideas from a linear system (considered in~\cite{HHK}) to a
nonlinear one (considered here) is neither evident nor trivial.
Indeed, holding the $\PT$-symmetry with nondissipative (or
weakly-dissipative) nonlinearity imposes additional (compared with
the linear case) constraints on the parameters of the applied
laser beams.

In the paraxial and the weakly guiding approximations
the equation governing the envelope of the probe field beam $\Omega_p$
 reads $2ik_p\frac{\partial \Omega_p}{\partial z}
+\frac{\partial^2 \Omega_p}{\partial
x^2}+k_p^2\chi_{p}\Omega_p=0$. Here $\chi_p$ is the probe-field
susceptibility. Outside the atomic cell it equals the effective
susceptibility of the Bragg cladding. Inside the  cell it is
given by $\chi_p=p_{eg}^2(N_1 \rho_{eg}^1+N_2
\rho_{eg}^2)/(\epsilon_0 \hbar\Omega_{p})$, where $N_{s}$
($s=1,2$) are densities and
$\rho_{eg}^s$ are coherences of the $s$-th atomic specie. The  coherences  $\rho_{eg}^s$ 
 can
be computed from the Bloch equations~\cite{fle}:
\bes\label{OBE}
\bea
& & i \dot{\rho}_{gg}^s=i\Gamma_{eg}\rho_{ee}^s-\bar{\Omega}_p\rho_{eg}^s+\Omega_p\bar{\rho}_{eg}^s,
\label{r11}\\
& & i\dot{\rho}_{aa}^s=i\Gamma_{ea}\rho_{ee}^s-\bar{\Omega}_c\rho_{ea}^s+\Omega_c\bar{\rho}_{ea}^s,
\label{r22}\\
& & i\dot{\rho}_{ee}^s =-i(\Gamma_{eg}+\Gamma_{ea})
\rho_{ee}^s+\bar{\Omega}_p\rho_{eg}^s-\Omega_p\bar{\rho}_{eg}^s
\nonumber\\
& & \qquad
+\bar{\Omega}_c\rho_{ea}^s-\Omega_c\bar{\rho}_{ea}^s,
\label{r33}\\
& &
i\dot{\rho}_{ag}^s=-d_{ag}^s\rho_{ag}^s+\Omega_p\bar{\rho}_{ea}^s-\bar{\Omega}_c\rho_{eg}^s,
\label{r21}
\\
& & i\dot{\rho}_{eg}^s
=-d_{eg}^s\rho_{eg}^s+\Omega_p(\rho_{ee}^s-\rho_{gg}^s)
-\Omega_c\rho_{ag}^s,
\label{r31}\\
& & i\dot{\rho}_{ea}^s
=-d_{ea}^s\rho_{ea}^s+\Omega_c(\rho_{ee}^s-\rho_{aa}^s)
-\Omega_p \bar{\rho}_{ag}^s.\label{r32}
\eea
\ees
In Eqs.~(\ref{OBE}) the overdots stand for the time derivatives, and we defined: 
$d_{ag}^s=-\delta_{s}+i\Gamma_{ag}/2$,
$d_{ea}^s=-\Delta_{s}+i(\Gamma_{eg}+\Gamma_{ea}+\Gamma_{ag})/2$,
and
$d_{eg}^s=-\delta_{s}-\Delta_{s}+i(\Gamma_{eg}+\Gamma_{ea})/2$
with $\Gamma _{jk}$ ($j,k=g,a,e$) being the spontaneous
emission decay rate from $|j,s\rangle$ to $|k,s\rangle$.

We are interested in the stationary states of (\ref{OBE}). Using the smallness of the intensity of
the probe field, i.e. considering $|\Omega_p/\Omega_c|\ll 1$, we look for a stationary solution in
the form
$\rho_{jk}^s=\sum_{m=0}^\infty\rho_{jk,m}^{s}$ where $j, k=g, a,
e$ and  $\rho_{jk,m}^{s}$ is of order of
$\left(\Omega_p/\Omega_c\right)^m$. Then, in the leading order the system (\ref{OBE}) is solved by
${\rho}_{gg,0}^{1}={\rho}_{aa,0}^{2}=1$ and
${\rho}_{ea,0}^{2}=-\Omega_c/d_{ea}^2$ with other leading elements
of the density matrix being zero.  Using the
conservation $\rho_{ee,m}^{s}=\delta_{0,m}-\rho_{gg,m}^{s}-\rho_{aa,m}^{s}$
[following from (\ref{r11})-(\ref{r33})],  in higher orders, we obtain the
recurrent relations
\bes\label{densityEq}
\bea \left(\begin{matrix}
d_{ag}^s &  \bar{\Omega}_c\\
\Omega_c &  d_{eg}^s
\end{matrix}\right)
\left(\begin{matrix}
\rho_{ag,m}^{s} \\
\rho_{eg,m}^{s}
\end{matrix}\right)=\Omega_p\left(\begin{matrix}
\bar{\rho}_{ea,m-1}^{s} \\
\rho_{ee,m-1}^{s}-\rho_{gg,m-1}^{s}
\end{matrix}\right), \label{densityEq1}\\
\left(\begin{matrix}
i\Gamma_{eg} & i\Gamma_{eg} & 0 & 0\\
i\Gamma_{ea} & i\Gamma_{ea} & \bar{\Omega}_c & -\Omega_c\\
\Omega_c & 2\Omega_c & d_{ea}^s & 0\\
-\bar{\Omega}_c & -2\bar{\Omega}_c & 0 & -\bar{d}_{ea}^s
\end{matrix}\right)  \left(\begin{matrix}
\rho_{gg,m}^{s} \\
\rho_{aa,m}^{s} \\
\rho_{ea,m}^{s} \\
\rho_{ae,m}^{s}
\end{matrix}\right)
=\mathbf{M}_{m-1}, \label{densityEq2}\eea
\ees
where
\be
\mathbf{M}_m=\left(\begin{matrix}
\Omega_p\bar{\rho}_{eg,m}^{s}-\bar{\Omega}_p\rho_{eg,m}^{s}
\\ 0 \\
-\Omega_p\bar{\rho}_{ag,m}^{s} \\
\bar{\Omega}_p\rho_{ag,m}^{s}
\end{matrix}\right). \nonumber
\ee

Having computed the coherence $\rho_{eg}^s$ up to the
third order of $|\Omega_p/\Omega_c|$, the probe-field susceptibility in
the atomic cell can be expressed as
$\chi_{p,1}+\chi_{p,3}|\Omega_p|^{2}$, where the
first- and third-order susceptibilities are
\bes
\bea \chi_{p,1}=\frac{p_{eg}^2}{\epsilon_0 \hbar}(N_1A_1+N_2A_2),
\label{chi1}
\\
\label{chi3}
\chi_{p,3}=\frac{p_{eg}^2}{\epsilon_0 \hbar
\Gamma_{ea}}\sum_{s=1}^2N_s\frac{d_{ag}^sC_s - B_s}{|\Omega_c|^2
-d_{ag}^sd_{eg}^s}, \label{chi3}
\eea\ees
with $A_1= -d_{ag}^1D_1$, $A_2=
-|\Omega_c|^2D_2/d_{eg}^2 $,
\bea
B_s=\frac{\Gamma_{ea} d_{ea}^s }{\Gamma_{eg}}
\mbox{Im}(A_s)
-|\Omega_c|^2 \mbox{Im}(D_s),
\nonumber\\
C_s=\mbox{Im}\left(
d_{ea}^sD_s-6\frac{\Gamma_{ea}}{\Gamma_{eg}}A_s+
\frac{|d_{ea}^s|^2}{|\Omega_c|^2}
\frac{\Gamma_{ea}}{\Gamma_{eg}}A_s\right), \nonumber
%
%
%
\nonumber\\
D_1=\frac{1}{d_{ag}^1d_{eg}^1-|\Omega_c|^2}, \quad
D_2=\frac{d_{eg}^2}{\bar{d}_{ea}^2(|\Omega_c|^2
-d_{ag}^2d_{eg}^2)}. \nonumber
\eea

 $\PT$-symmetric   spatial distribution of the susceptibility,
 $\chi_{p,1}(x)=\bar{\chi}_{p,1}(-x)$, is achieved with help of the far-detuned laser field $E_{\rm
S}(x)\cos(\omega_S t)$~\cite{HHK}, referred to as the Stark field. This field induces
the shifts of the levels $|j,s\rangle$: $\Delta E_{j,S}(x)=-
\alpha_{j} E_{\rm S}^2(x)/4$ (here $\alpha_{j}$ is the
scalar polarizability), and as a consequence the
$x$-dependence of the control field $\Omega_{c}=\Omega_{c}(x)$.
 Since within the required accuracy $\alpha_{g}\approx\alpha_{a}$, the difference of the
Stark shifts between the ground-state sublevels is negligible, i.e.
$\delta_{s}$ is not affected by $E_S$, while
$\Delta_{s}(x)=\Delta_{s}- (\alpha_{e}-\alpha_{g})E_{\rm
S}^2(x)/4\hbar$, $\Delta_{s}$ being here the one-photon detuning of the $s$-th isotope in the absence of the Stark field.
We notice that the characteristic scale of the  $\Delta_{s}(x)$
modulation is of order of the Stark field wavelength $\lambda_{\rm
S}$. For a typical magnitude of the control field  $E_c\sim 1$~V/cm
($\Omega_c\sim 10^7$~Hz), the required amplitude of the Stark
field is $E_{\rm S}\sim 10^3$~V/cm. Being focused
into a spot of a millimeter diameter, this requires laser powers
of order of 10~W.

As a case example, we explore a mixture of
$^{85}$Rb ($s=1$) and $^{87}$Rb ($s=2$) isotopes, and assign
$|g,s\rangle=|5S_{1/2}, F=1\rangle$, $|a,s\rangle=|5S_{1/2},
F=2\rangle$, and $|e,s\rangle=|5P_{1/2}, F=1\rangle$. The densities for each isotope are
$N_1\approx 6\times10^{14}$ cm$^{-3}$ and
$N_2\approx 1.92\times10^{15}$~cm$^{-3}$. The coherence decay
rates are estimated as
$\Gamma_{eg}\approx\Gamma_{ea}\approx\pi\times5.75$~MHz and
$\Gamma_{ag}\approx10^{-3} \,\Gamma_{eg}$. With sufficiently high
accuracy $\alpha_{e}-\alpha_{g}=2\pi\hbar\times 0.1223$
Hz(cm/V)$^{2}$ and $p_{eg}=2.5377\times10^{-27}$ C
cm~\cite{Steck}. The two-photon detunings are chosen
as $\delta_{1}=-0.18\,\Gamma_{eg}$ and $\delta_{2}=1.63\,\Gamma_{eg}$.
Following the algorithm of~\cite{HHK}, we compute that the $\PT$-symmetric
profile of the susceptibility  (with parabolic
real part and  linear imaginary part) can be achieved by using the control
and Stark fields with the forms
\bes
\label{field}
\bea \Omega_c(\xi) =
(2.5+0.025\,\xi+2.4741\times10^{-4}\xi^2)\Gamma_{eg},\\
E_{\rm S}(\xi) =(1.9394-0.2542\,\xi)E_0,
\eea
\ees
where $\xi=k_S x$ ($k_S=2\pi/\lambda_S$) and $E_0=5.0\times 10^3$
V/cm.

At this stage it is important to emphasize that Eqs.~(\ref{field}) provide accurate  $\PT$ symmetry only for sufficiently small $\xi$, while for large $\xi$ significant deviations are observed (see the discussion in~\cite{HHK}). This imposes a constraint on the choice of the size of the atomic cell, which is needed to ``cut'' the undesirable deviations form the $\PT$ symmetry at large $\xi$. For the sake of definiteness, we impose the size of the cell to be $\lambda_S/\pi\approx 0.04\,$mm where $\lambda_S\approx 0.13\,$mm is the typical wavelength of the applied Stark field, which in the dimensionless units correspond to the cell occupying the domain $|\xi|<1$ [Fig.~\ref{fig1}(b)].

With the above parameters of the atomic cell and choosing the effective
susceptibility for the Bragg cladding $\chi_{\rm
clad}\approx-0.64046$~\cite{Dowling,Bend} ($n_{\rm clad} \approx
0.5996$), the first-order probe-field susceptibility
acquires the form
\be\label{chi_p1} \chi_{p,1}(\xi)\approx \left\{\begin{array}{ll}
\tilde{\chi}_{0}+ i
\tilde{\chi}_{1}\xi+\tilde{\chi}_{2}\xi^2,&
|\xi|\leq 1, \\
\chi_{\rm clad}, & |\xi|>1,
\end{array}
\right.
\ee
where $\tilde{\chi}_0\approx-0.6398\approx\chi_{\rm
clad}+0.0007$, $\tilde{\chi}_{1}\approx -3.2380\times
10^{-4}$, and $\tilde{\chi}_{2}\approx -3.6688\times 10^{-4}$.
In Fig.~\ref{fig1}(b) we show 
real 
and imaginary 
parts of the linear
refractive index of the defect. In order  to control the
accuracy of $\PT$ symmetry, we calculated the error
function~\cite{HHK} $\nu(\xi)\equiv
\chi_{p,1}(\xi)-\bar{\chi}_{p,1}(-\xi)$ for $|\xi|\leq1$. Real
and imaginary parts of $\nu$ 
are $\sim
10^{-7}$ and $\sim 10^{-10}$, respectively.

Another important observation is in order here. In an infinite medium    the constant part of the refractive index did not play any significant role~\cite{HHK}. In the case of a defect mode, however, it becomes relevant for determining the guiding regime. In particular, it was mentioned above that we are dealing   with the weak guidance, which is ensured by the fact that $|\tilde{\chi}_0- \chi_{\rm clad}| \sim 10^{-3}\chi_{\rm clad}$. This   gives the   order of small parameter $|\Omega_p/\Omega_c|^2\sim 10^{-3}$ and hence defines the accuracy of the expansion.

Turning to the third-order probe-field susceptibility, we compute it using Eq.~(\ref{chi3}). For the above given parameters  inside the atomic cell ($|\xi|\leq1$) we find
$\tilde{\chi}_{p,3}\approx -0.1294\times 10^{-14}$. Generally speaking,
the real and imaginary parts of $\tilde{\chi}_{p,3}$ depend on the spatial coordinates and also violate the $\PT$ symmetry. These effects, however, are $10^{-2}$ times
smaller than the leading order, i.e. are beyond the accepted accuracy.
Now, from the relation $n_2 = p_{eg}^2\tilde{\chi}_{p,3}/(2\hbar^2n_0)$,
we estimate $n_{2}\approx-0.6244$ cm$^2$V$^2$ for the
defect. This is about $10^{16}$ larger than that measured for
passive optical materials. Such
enhancement 
of the 
nonlinearity 
 occurs the
existence of two nearly resonant Raman transitions.


Resuming the above parameters, 
the probe field in the cell is described by the nonlinear Schr\"{o}dinger equation  with a
$\PT$-symmetric linear potential, while outside the   cell the system is described by the linear equation:
\begin{subequations}
\label{MAX}
\begin{eqnarray}\label{MAX2}
 i\frac{\partial u}{\partial \zeta}+ \frac{\partial^2u}{\partial
\xi^2} - (iV_1\xi+V_2\xi^2) u -|u|^2u=0, \quad |\xi|< 1,
\\
\label{MAX3} i\frac{\partial u}{\partial \zeta} +\frac{\partial^2
u}{\partial \xi^2} -V_0u=0 \quad |\xi|\geq 1.
\end{eqnarray}
\end{subequations}
Here $u=\Omega_pe^{-ik_p\tilde{\chi}_0z/2}/\Omega_{0}$ is the dimensionless field, whose normalization constant $\Omega_{0}$ is chosen to scale out the nonlinear coefficient in (\ref{MAX2}),  $\zeta= (k_S^2/2k_p)z$, $V_0= (k_p^2/k_S^2)
(\tilde{\chi}_{0}-\chi_{\rm clad})$, $V_{1,2}=
- (k_p^2/k_S^2)\tilde{\chi}_{1,2}$. For the mixture of the rubidium isotopes and $\lambda_S=0.13\,$mm (thus $ k_p/k_S\approx10^2$) we have $\Omega_{0}=2.8\times10^5\,$s$^{-1}$ $V_0\approx 7.1$, $V_1\approx 3.2380$, and  $V_2\approx 3.6688$.

\begin{figure}
\centering
\includegraphics[width=\columnwidth]{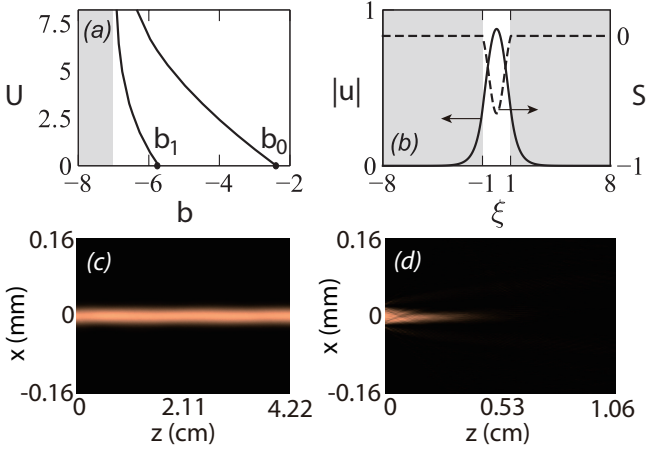}
\caption{(Color online) (a):
Two families of nonlinear modes bifurcating from the eigenvalues
$b_{0,1}$ (see the text). The shaded domain corresponds to propagation constants belonging to the continuous spectrum. (b): The amplitude
$|u|$ (solid line) and the current $S$ (dashed line) for the
nonlinear mode with $b=-3$. The shaded domain corresponds to the
cladding. (c), (d): Evolution of the nonlinear
mode intensity $|u|^2$ for the $\PT$-symmetric (c) and
non-$\PT$-symmetric (d) refractive indexes.} \label{fig2}
\end{figure}

Turning now  to the properties of the  defect modes, we first notice that  nonlinear modes in the parabolic $\PT$-symmetric potential were described in~\cite{ZK}. Here, however, we are dealing with a cut parabolic profile, which naturally affects properties of the system.  Indeed, let us consider the linearized version of system   (\ref{MAX}). Without the nonlinear term in (\ref{MAX2}), the stationary guided modes  $u=e^{ib\zeta}w(\xi)$  are determined by the eigenvalue problem ${\cal L}w =
b w$, where $b$ is the propagation constant and  the $\PT$-symmetric operator ${\cal L}$ is defined as
${\cal L} = \frac{d^2}{d\xi^2} - V(\xi)$ with
$V(\xi)=iV_1\xi+V_2\xi^2$ for $|\xi|<1$ and $V(\xi)=V_0$
otherwise. For the above parameters, we have found numerically that ${\cal L}$ has exactly two isolated
eigenvalues $b_0 \approx -2.4$ and $b_1 \approx -5.8$ [see Fig.~\ref{fig2}(a)] and the
continuous spectrum situated on the real axis. Reality of the spectrum of   ${\cal L}$ is an evidence of the unbroken $\PT$ symmetry.

The existence of the two bound states is supported by the given positive value of $V_0$. Adjusting the parameters, one can change $V(\xi)$ and hence    the properties of the defect modes. For example, decrease
(increase) of $\chi_{\rm {clad}}$ results in sequential appearance
(disappearance) of isolated real eigenvalues $b$ meaning change of
the number of  guided linear modes. In particular, for $\chi_{\rm{clad}}> -0.64$ no
linear guided modes exists.
Increase of the imaginary part of the potential eventually results in the   $\PT$ symmetry breaking  (for the chosen parameters,    $\tilde{\chi}_{1}$ must be about doubled for the $\PT$ symmetry breaking to occur).

Passing to the nonlinear case,  each eigenvalue (i.e. $b_0$, $b_1$, etc.) gives birth to a
family of nonlinear modes, as shown in
Fig.~\ref{fig2}(a) on the plane ($U,b$), where
$U = \int_{-\infty}^\infty|u|^2d\xi$ is the total energy flow. In the  linear limit,
$U\to 0$, the
propagation constants approach $b_{0,1}$. With the
decrease of $b$ the total energy flow monotonously grows. A
typical profile of a defect mode is shown in Fig.~\ref{fig2}(b), where we  also plot the ``current'' $S=\frac{i}{2}(u\frac{\partial
\bar{u}}{\partial \xi}-\bar{u}\frac{\partial u}{\partial \xi})$,
which is  associated with the
power-flow density (i.e. with the Poynting vector) in the transverse
direction. This current arises from the nontrivial phase structure
of the nonlinear    modes. The mode shown in   Fig.~\ref{fig2}(b) is  well localized inside the
atomic cell:   $94\%$ of its energy  is confined in the cell.

To examine stability of the  modes
in a practical system, we add small (of order of 5\%) random
perturbations to both amplitude and phase of the stationary
solution shown in Fig.~\ref{fig2}(b) and evolve it according to Eqs.~(\ref{MAX}). The mode displays robust evolution shown   in Fig.~\ref{fig2}(c). For comparison, in
Fig.~\ref{fig2}(d), we repeat the evolution of the same input beam after decreasing $N_2$
by 5\% without changing other parameters. In this case, the system is not $\PT$ symmetric any more [specifically, now one has  $ \chi_{p,1}\approx-0.5860-(3.3933\xi^2+3.9006\xi+i4.9495-i3.2402\xi)\times10^{-4}$]. As expected, we observe decay of the  mode which is absorbed at relatively short distances.


To conclude, we have proposed a 
scheme of creating strongly nonlinear tunable defects in photonic crystals. Such defects consist of an atomic cell filled in with a mixture of isotopes of lambda atoms subjected to the control and to the  Stark fields.
By proper adjusting the parameters of the control field the defect can be made $\PT$-symmetric and in this case it supports families of stable defect modes of the probe field. Due to flexibility, the model allows one to study the effect of nonlinearity on the bound state number and to design defects with focusing nonlinearities which in their turn may support the quasi-bound states~\cite{Moiseyev}. The generation power of such defect modes can be reduced to the nanowatt range due to resonant enhancement of Kerr nonlinearity.


{\bf Acknowledgments} The work was supported by NSF-China under
Grant Nos. 11174080 and 11105052, by the Program of Introducing
Talents of Discipline to Universities under Grant No. B12024, and
by the FCT (Portugal) grants PTDC/FIS-OPT/1918/2012 and 
PEst-OE/FIS/UI0618/2011. 

\end{document}